\documentclass[twocolumn]{aastex63}

\usepackage{xcolor}
\usepackage{natbib}
\usepackage{graphicx}
\usepackage{amssymb}
\usepackage{multirow}
\usepackage{wrapfig}
\usepackage{enumitem}
\usepackage[varg]{txfonts}
\usepackage[toc,title,page]{appendix}
\usepackage{csquotes}

\def\baselinestretch{1}
\bibpunct{(}{)}{;}{a}{}{,}

\usepackage{hyperref}

\newcommand*{\z}{$|Z|$}
\newcommand*{\I}{$(2I_3)^{1/2}$}

\newcommand*{\kpckms}{kpc~km s$^{-1}$}

\newcommand*{\en}{$\times 10^5$ km$^2$~s$^{-2}$}
\newcommand*{\tho}{$\theta_{\rm orb}$}

\definecolor{tropicalrainforest}{rgb}{0.0, 0.46, 0.37}
\definecolor{darkmagenta}{rgb}{0.55, 0.0, 0.55}

\hypersetup{
    bookmarks=true,         % show bookmarks bar?
    unicode=false,          % non-Latin characters in Acrobat’s bookmarks
    pdftoolbar=true,        % show Acrobat’s toolbar?
    pdfmenubar=true,        % show Acrobat’s menu?
    pdffitwindow=true,     % window fit to page when opened
    pdfstartview={FitH},    % fits the width of the page to the window
    pdftitle={},    % title
    pdfauthor={Beers},     % author
    pdfsubject={Astronomy},   % subject of the document
    pdfcreator={dvipdf},   % creator of the document
    pdfproducer={dvipdf}, % producer of the document
    pdfkeywords={metal-poor stars},% list of keywords
    pdfnewwindow=true,      % links in new window
    colorlinks=true,       % false: boxed links; true: colored links
    linkcolor=red,          % color of internal links
    citecolor=blue,        % color of links to bibliography
    filecolor=magenta,      % color of file links
    urlcolor=cyan,           % color of external links
    breaklinks=true,
    linktocpage
}

\begin{document}

\title{Understanding the early stages of galaxy formation using very metal-poor stars from the Hamburg/ESO survey}
%\title{The nature of the Milky Way's stellar halo selected from the %three integrals of motion space} 
% Tracking down the first stages of galactic evolution: very metal-poor stars in the Hamburg/ESO survey

\author{Daniela Carollo}

\affiliation{INAF - Osservatorio Astronomico di Trieste, I-34143 Trieste, Italy}

\author{Norbert Christlieb}
\affiliation{Zentrum f\"ur Astronomie der Universit\"at Heidelberg, Landessternwarte, K\"onigstuhl 12, 69117 Heidelberg, Germany}

\author{Patricia B. Tissera}
\affiliation{Institute of Astronomy, Pontificia Universidad Cat\'olica de
Chile, Av. Vicu\~na Mackenna 4860, 782-0436 Macul, Santiago, Chile.} 
\affiliation{Centro de Astro-Ingenier\'ia, Pontificia Universidad Cat\'olica de
Chile, Av. Vicu\~na Mackenna 4860, 782-0436 Macul, Santiago, Chile.}
\affiliation{N\'ucleo Milenio ERIS, Av. Vicu\~na Mackenna 4860, 782-0436 Macul, Santiago, Chile.}
\author{Emanuel Sillero}
\affiliation{Institute of Astronomy, Pontificia Universidad Cat\'olica de
Chile, Av. Vicu\~na Mackenna 4860, 782-0436 Macul, Santiago, Chile.}

\begin{abstract}
\noindent
We explore the chemo-dynamical properties of a sample of very metal-poor (VMP) stars selected from the Hamburg/ESO survey, matched with Gaia EDR3, in the phase-space identified by the three integrals of motion ($L_z$, $E$, $I_3$). Disk and halo orbits are separated by using the criteria defined in \citet{carollo2021}. We found 26 stars with $\mbox{[Fe/H]} \leq -2.5$ possessing disk kinematics, of which 14 are extremely metal-poor. At these metallicities, the number of stars with disk kinematics is three times its retrograde counterpart. In the same range of metallicity we also identified 37 halo stars most tightly bound to the gravitational potential of the progenitor halo. The origin of these stars are investigated by comparing the observational results with simulated galaxies from the Aquarius Project and the IllustrisTNG simulations. We found two mechanisms of formation of VMP stars with disk kinematics: accretion from early satellites (which is dominant), and {\it in-situ} formation. These stars are very old, with ages $>$ 12.5 Gyr ($z >$ 5), and they are $\alpha$-enriched. Accretion and {\it in-situ} formation are also found for the retrograde counterparts with being accretion also the dominant mode. Contributing accreted satellites have stellar masses in the range $[10^{6}-10^9]$ M$_{\sun}$, and are very gas-rich. The most bound halo stars are the oldest detected with a median age of $\sim$ 13.3 Gyr ($z \sim$ 11), and $\alpha$-enriched. Our finding clearly show that very old, very metal-poor stars store important information on the first stages of assembly of our Galaxy and its halo. \\

\end{abstract}


\section{Introduction}
\noindent
Very metal-poor (VMP; [Fe/H] $< -$2.0) stars are of great interest in near-field cosmology, because they preserve a fossil record of the nucleosynthesis products of the first generations of stars that formed shortly after the Big Bang. Hence they offer a plethora of information on the early stages of galaxy formation and the chemical evolution of galactic halos. These stars are mainly located in the Milky Way (MW) halo, which, in terms of global properties, comprises “at least” two stellar populations, inner- and outer-halo, with
different kinematics, spatial distribution, and chemical composition (\citealt{carollo2007,carollo2010}; see also \citealt{beers2012}).\\

\noindent
In a recent investigation, \citet{carollo2021} (hereafter CC21) explored the general properties, or {\it coarse-grained} phase-space distribution, of the Milky Way's halo system by using a large sample of Sloan Digital Sky Survey (SDSS)-SEGUE DR7 and Apache Point Observatory Galaxy Evolution Experiment (APOGEE) DR16 catalogs \citep{ahumada2020}, matched with Gaia DR2. The phase-space is defined by the three integrals of motion ($L_z$,$E$,$I_3$), where $L_z$ is the vertical angular momentum, $E$ is the total energy, and $I_3$ is third integral of motion, which is analytically defined in a St\"{a}ckel-form gravitational potential. Besides the definition of a new method to select halo stars without introducing a bias associated with high-velocity cuts, this analysis shows that the inner halo stellar population includes: 1) the Gaia Enceladus Sausage structure (GES),  (\citealt{helmi2018}, \citealt{belokurov2018}; see also \citealt{nissen2010} and \citealt{haywood2018} for earlier hints of the existence of this accretion) which dominates this component at a metallicity of [Fe/H] $\sim -1.5$; 2) low-energy stars (E $<-1.5$ \en) at [Fe/H] $< -1$; and, 3) a significant number of metal-poor prograde stars. It was shown that the GES structure exhibit a metallicity distribution function (MDF) with a metal-weak tail much steeper than the average MW (\citealt{bonifacio2021}, and reference therein). This implies that it is unlikely to find VMP stars in this structure.

\noindent
CC21 also confirms that the outer halo component is dominated by VMP stars (peak at [Fe/H] $\sim -$2.2), and possesses predominantly retrograde motion. Such a coarse-grained structure is likely made of a superposition of several substructures, or “fine-grained” elements, such as Sequoia \citep{myeong2019}, Arjuna and I’itoi \citep{naidu2020}, and the dynamically tagged groups (DTG) identified in
\citet{yuan2020}. Inner halo stars are still present in the VMP outer halo but they do not represent a significant fraction (see Figure 10 in CC21). \\

%It has been shown that the metal-weak tail of the 5metallicity distribution function (MDF) of the GE debris 5is much steeper than the average MW %(\citealt{bonifacio2021}, and reference therein). This 5implies that is unlikely to find VMP stars in this 5structure.\\
\noindent
CC21 showed that halo stars have values of the third integral of motion parameter, \I $>$ 1000 \kpckms, which corresponds to a maximum orbital angle of $\theta_{orb} >$ 15--20$^{\circ}$. On the contrary disk(s) stars possess low orbital angles from the Galactic plane, below 15--20$^{\circ}$. Moreover, the fraction of very metal-poor stars with [Fe/H] $<$ $-$2.2, comprising the outer halo, relative to more metal-rich halo stars ($-$2 $<$ [Fe/H] $<$ $-$1.4), starts to dominate at \I $\sim$ 2000 \kpckms, corresponding to $\theta_{orb} \sim$ 50$^{\circ}$ (Z$_{\rm max} \sim 10$\,kpc), at the Solar radius. 

\noindent
Based on these findings, one would expect to find VMP stars primarily at large orbital angles on a prograde or retrograde motion. While this is certainly the case, recent works report the existence of VMP stars possessing angular momenta and orbits typical of the MW's disk stellar populations. In these investigations the separation between disk and halo stars was done in action space \citep{sestito2019, sestito2020}, or by adopting high velocity cuts \citep{dimatteo2020,venn2020}. 

\noindent
In CC21 it was clearly shown that the high-velocity cuts method is highly kinematically biased against those halo stars possessing low relative velocity with respect to the Local Standard of Rest (LSR), which likely have low orbital eccentricities, and the adoption of this method can cause halo stars to be misclassified as disk, or vice versa.

\noindent
The origin of VMP stars with disk kinematics is not clear. \citet{sestito2021} envisage two possible scenarios. In the first, the proto-galaxy, and its proto-disk, was assembled via a chaotic process of accretion of systems in the mass range of 10$^{5}$--10$^{9}$ M$_{\odot}$, which can deposit stars in prograde or retrograde motion. The second scenario involves a later merger event whose stars are stripped and assimilated in the disk component, depending on their orbital parameters.

\noindent
CC21 discuss the possibility that some VMP stars with disk kinematics may have been formed in-situ, from an early infalling pristine gas that would have been settled into an
equatorial plane of a progenitor dark halo, in the presence of an
initial angular momentum (e.g., \citealt{katz1991}). A possible in-situ origin was also envisaged by CC21 for the most bound stars in the MW's halo, i.e. stars with the lowest binding energy, i.e. $E < -$1.5 \en\footnote{In a St\"{a}ckel potential}.

\noindent
In this work we use a sample of VMP stars from the Hamburg/ESO Survey \citep[HES;][]{reimers1990} matched with Gaia EDR3 \citep{gaia2020}, and analyze their chemo-dynamical properties in the integrals of motion phase-space. We show that a fraction of them, including EMP and UMP stars, possess disk kinematics. Insights on the possible origin of these peculiar stars is done by using two sets of simulations: the Aquarius suite \citep{scannapieco2009} from which we select the Aq-C halo, because it hosts a disc-dominated galaxy and reproduces a diversity of observations of the MW stellar halo \citep{tissera2017,fernandez-alvar2018, santos-santos2020}, and the TNG50 simulations (\citealt{nelson2019b,pillepich2019a}; see \citealt{nelson2019a} for the IllustrisTNG data relase), from which we select four MW's analogs. We also analyze the properties of the most-bound VMP stars in the halo system, and discuss their origin.

\noindent
The paper is organised as follows. Section 2 describes the data sets, the derivation of kinematic and dynamic parameters, and the adopted mass model. In Section 3 we analyze the distribution of stars in phase-space and its dependence on metallicity, [Fe/H]. In the same section we characterize the VMP stars with disk kinematics and describe their properties, as well as the most bound stars in the halo system. Section 4 present the analysis of simulated halos and comparison with the observational results, a discussion about the origin of the two dynamically informative groups of stars, and the 
implications for the early stages of galaxy formation. Summary and conclusions are in Section 5. \\

\section{Data: Hamburg/ESO Survey and {\it Gaia} eDR3}

\noindent
The Hamburg/ESO Survey \citep[HES;][]{reimers1990} is an objective-prism survey that was carried out with the 1\,m ESO Schmidt telescope. Originally conceived as a survey for bright (i.e., $B \lesssim 18$) quasars, it covers the southern high Galactic latitude sky \citep{wisotzki2000}. However, the spectral resolution of $\Delta\lambda \approx 
10$\,{\AA} at the Ca~II~K line made it possible to also select stellar objects efficiently, such as white dwarfs of 
spectral type DA \citep{christlieb2001}, or field horizontal-branch stars \citep{christlieb2005}. 
\citet{christlieb2008} developed techniques for automated selection of candidate metal-poor stars, and applied that 
selection to a nominal survey area of 8853\,deg$^2$. The selection was based on the $B-V$ color and strength of the 
Ca~II~K line, both measured in the HES objective-prism spectra. The selection resulted in 20,271 stars in the magnitude range $10\lesssim B\lesssim 18$. 

\noindent
Spectroscopic follow-up observations of most of the 4519 HES stars used in this paper were carried out during the period April 2000 to July 2007 with the five telescope/instrument combinations listed in Table~2 of \citet{schorck2009}, and by the observers mentioned there. The spectra cover at least the wavelength region 390--440\,nm, and they typically have a resolving power of $R=\lambda/\Delta\lambda =2000$. The vast majority of the spectra have a signal-to-noise ratio ($S/N$) of more than 20 per pixel in the continuum near the Ca~II~H and K lines.

\noindent
The wavelength calibration of the spectra was performed by measuring the positions of emission lines in calibration lamp spectra, and deriving a polynomial solution of the wavelength as a function of the $x$-position on the CCD, where $x$ is the dispersion direction. However, since the spectra were originally not intended to be used for a kinematic analysis, but just for measuring the metallicities of the stars, wavelength calibration spectra were usually only acquired in the beginning and at the end of the night when the telescope was in its parking position, and not while the telescope was pointing towards the targets. We can therefore not exclude that radial velocity uncertainties of the order of 5--10\,km/s were introduced by spectrograph flexure. 

\noindent
The geocentric radial velocities were measured by fitting Gaussian profiles to a few strong lines clearly visible in the spectra, resulting in a precision of typically 10-15\,km/s. From the information stored in the FITS headers of the raw data files, barycentric corrections were computed, and these were then applied to the radial velocity measurements.

\noindent
The metallicities (i.e., [Fe/H]) were determined with an updated version of the method of \citet{beers1999}, which is based on measurements of the line index KP for the Ca~II~K line, and the HP2 index for the H$\delta$ line. We removed 242 stars with $\mathrm{GP} > 6$ from the sample, where GP is the line index of the G band of CH. This was done to avoid systematically underestimating [Fe/H] in stars showing a strong G band in their spectra, leading to too low estimates of the continuum level \citep{Cohenetal:2005}. The typical uncertainty of the [Fe/H] measurements within the sample of stars with $\mathrm{GP} < 6$ is $\sim 0.3$\,dex, as was determined from a comparison with Fe abundances 
based on high-resolution spectra \citep{schorck2009}.

\noindent
Table \ref{Tab:SampleStatistics}, first column, reports the number of stars in the low metallicity range for the original sample, starting from [Fe/H] = $-$2. 

\begin{deluxetable}{lcc}
\tablecolumns{3}
\tablewidth{0pt}
\label{Tab:SampleStatistics}
\tablecaption{Number of stars in the low metallicity ranges}
\tablehead{
\colhead{} &
\colhead{Original Sample} &
\colhead{Full Sample}
}
\startdata
$-$2.5 $<$ [Fe/H] $\leq -$2.0   & 1174  &  1067 \\
$-$3.0 $<$ [Fe/H] $\leq -$2.5   & 574  &  538   \\
$-$3.5 $<$ [Fe/H] $\leq -$3.0   & 208  &  201   \\
$-$4.0 $<$ [Fe/H] $\leq -$3.5   & 27  &  26     \\
\hspace{0.85cm} $ {\rm [Fe/H]} \leq -$4.0 & 11  &  11\\
\hline
VMP \hspace{0.1cm} ([Fe/H] $< -$2) & 1976 & 1826\\
EMP \hspace{0.1cm} ([Fe/H] $< -$3) & 235 & 228 \\
UMP \hspace{0.1cm} ([Fe/H] $\leq -$4) & 10 & 10\\
\hline
\hline
\enddata
%\tablecomments{}
\end{deluxetable}

%\begin{deluxetable}{lcc}
%\tablecolumns{3}
%\tablewidth{0pt}
%\label{Tab:SampleStatistics}
%\tablecaption{Number of stars in the low metallicity ranges %UPDATE NUMBERS}
%\tablehead{
%\colhead{} &
%\colhead{Original Sample} &
%\colhead{Full Sample}
%}
%\startdata
%$-$2.5 $<$ [Fe/H] $< -$2.0   & 1187  &  1080 \\
%$-$3.0 $<$ [Fe/H] $< -$2.5   & 579  &  540   \\
%$-$3.5 $<$ [Fe/H] $< -$3.0   & 252  &  244   \\
%$-$4.0 $<$ [Fe/H] $< -$3.5   & 41  &  39     \\
%\hspace{0.85cm} $ {\rm [Fe/H]} <-$4.0 & 19  &  19\\
%\hline
%VMP \hspace{0.1cm} ([Fe/H] $< -$2) & 1766 & 1620\\
%EMP \hspace{0.1cm} ([Fe/H] $< -$3) & 293 & 283 \\
%UMP \hspace{0.1cm} ([Fe/H] $< -$4) & 19 & 19\\
%\hline
%\hline
%\enddata
%\tablecomments{}
%\end{deluxetable}

\noindent
This sample was cross-matched with the Gaia EDR3 \citep{gaia2020} database, using the CDS (Centre de Donnes Astronomiques de Strasbourg) X-Match service, and adopting a search radius of 1.5". The match provided positions, trigonometric parallaxes, and proper motions, for 92\,\% of the HES stars (i.e., 3906 out of 4277 stars). 

\noindent
We then divided the cross-matched sample into three sub-samples: 1) stars with relative parallax errors of $\sigma_{\pi}$/$\pi$ $<$ 0.2 (3013 stars), 2) stars with relative parallax errors of $\sigma_{\pi}$/$\pi$ $>$ 0.2 (840 stars), and 3) stars with negative parallax, $\pi <$ 0 (53 stars). This separation was done mainly to keep in the analysis a higher number of very metal-poor stars (i.e., stars with $\mathrm{[Fe/H]}< -2$; \citealt{beers2005}). In case of the first sub-sample, distances are derived using the relation $d$ = 1/$\pi$, and adopting a parallax zero-point offset of $\delta_{\pi}$ = $-$0.017 mas \citep{lindegren2020}. For the remaining sub-samples, we adopted the photogeometric distances as described in \citet{bailer-jones2021}. 

\noindent
Hereafter, we refer to the sample obtained by combining the three aforementioned datasets as "the full sample",  and the resulting number of stars in the low metallicity intervals are shown in the second column of Table \ref{Tab:SampleStatistics}.

%In sub-sample 1, the number of stars with metallicity %$-$2.5 $<$ [Fe/H] $< -$2, in $-$3 $<$ [Fe/H] $< -$2.5, %$-$3.5 $<$ [Fe/H] $< -$3, and [Fe/H] $< -$4, are 790, %360, 159, 27 and 16, respectively. 
%$-$2.5 $<$ [Fe/H] $< -$2, $-$3 $<$ [Fe/H] $< -$2.5, %$-$3.5 $<$ [Fe/H] $< -$3, $-$4 $<$ [Fe/H] $< -$3.5, and %[Fe/H] $< -$4, become, 1080, 540, 244, 39, and 19, %respectively. \\

\noindent
The average proper motion uncertainty in this sample is less than 0.2 mas yr$^{-1}$. The full space and orbital motion are derived by combining the observables obtained from {\it Gaia} EDR3, i.e., positions, distances, and proper motions ($\alpha$, $\delta$, $\pi$, $\mu_{\alpha}$, $\mu_{\delta}$), with the radial velocities measured in the moderate-resolution spectra.

\noindent
The velocities calculated in the LSR, assumed to be rotating at 220\,km\,s$^{-1}$, are referred to as $(U, V, W)$, which are corrected for the motion of the Sun by adopting the values
($U, V, W$) = ($-$9,12,7) km s$^{-1}$ (\citealt{mihalas1981})\footnote{More recent values of the LSR and Solar
motion are available, but we adopt these values for consistency with the analysis of CC21.}. The velocity
component $U$ is taken to be positive in the direction toward the Galactic anti-centre, the $V$ component
is positive in the direction toward Galactic rotation, and the $W$ component is positive toward the north Galactic 
pole.

\noindent
The adopted axisymmetric Galactic gravitational potential is of St\"{a}ckel type, which consists of a highly flattened disk and an oblate dark halo (\citealt{dezeeuw1986}; CC21; and references therein). In that model, we calculate the three isolating integrals of motion ($L_z$,$E$,$I_3$), where, $L_z$ is the angular-momentum component parallel to the $z$ axis, $E$ is the orbital energy , and $I_3$ is the third integral of motion. Other orbital parameters, such as r$_{\mathrm{\footnotesize apo}}$ (apocentric galactocentric radius), r$_{\mathrm{\footnotesize peri}}$ (pericentric galactocentric radius), and eccentricity, defined as (r$_{\rm apo}$ - r$_{\rm peri}$)/(r$_{\rm apo}$ + r$_{\rm peri}$), are also determined. Average errors on the phase-space parameters are 100 \kpckms for $L_z$, $0.2 \times 10^4$ km$^{2}$~s$^{-2}$ for $E$, and $\sigma_{I_3}$/$I_3$ $\sim$ 0.1\footnote{The uncertainties of the derived orbital parameters due to the observational errors have been estimated through a Monte Carlo simulation (100 realizations for each star).}. In case of $r_{\mathrm{\footnotesize apo}}$, $r_{\mathrm{\footnotesize peri}}$, and the eccentricity, the estimated 1$\sigma$ errors are 1.1\,kpc, 2.2\,kpc, and 0.12, respectively. 

\begin{figure}[t!]
\centering
%\vspace{-4cm}
%\hspace{3cm}
\includegraphics[width=80mm]{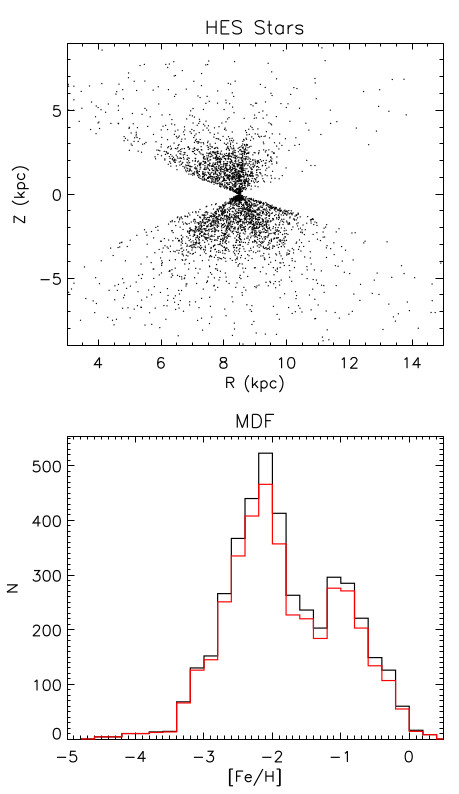}
\caption{Top-panel: Distribution of HES stars in the (Z,R) plane. Bottom-panel: Metallicity distribution function of the original HES sample (black solid line), and the adopted HES sample (red solid line). 
}
\label{fig:rz}
\end{figure}

\noindent
The top panel of Figure~\ref{fig:rz} shows the spatial distribution in the (R,Z) coordinates for the full sample. As expected for the HES data, the area below the Galactic plane is better represented than the northern Galaxy. Stars whose heliocentric distances were determined as d = 1/$\pi$ reach vertical distances of up to 5 kpc, while stars for which photogeometric distances were adopted reach up to $\rm z$ $\sim$ 8 kpc. The bottom panel represents the MDF for the original sample (black; 4277 stars), and the full sample (red; 3906 stars). It can be noted that in the process of matching with Gaia EDR3 data, we lost some very metal poor stars ($\sim$ 7\%), but we were able to keep almost all the extremely metal-poor stars (i.e., the stars at $\mathrm{[Fe/H]} < -3$).

%%% Fig. 1 %%%%%%%%%%%%%%%%%%%%%%%%%%
%%%%%%%%%%%%%%%%%%%%%%%%%%%%%%%%%%%%%%

%%% Fig. 2 %%%%%%%%%%%%%%%%%%%%%%%%%%%
\begin{figure*}[hbt!]
\centering
\includegraphics[width=\textwidth]{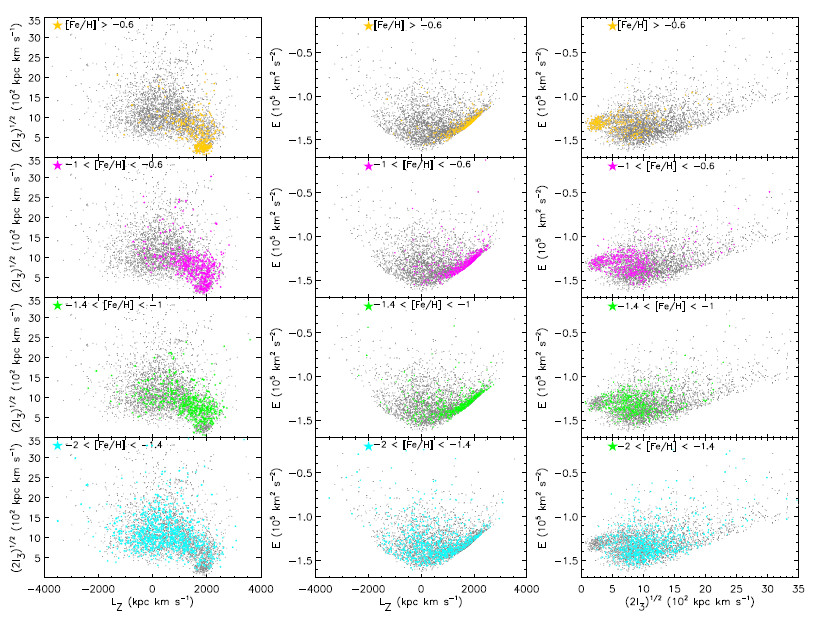}
\caption{
Distributions of the HES stars in the phase-space $(E, L_z, I_3)$ shown in terms of three diagrams,
\I \ vs. $L_z$ (left), $E$ vs. $L_z$ (middle), and $E$ vs. \I \ diagrams (right panel). In this figure,
four metallicity ranges are shown, [Fe/H]$>-0.6$ (top; ochre), $-1<$[Fe/H]$<-0.6$ (second panel; fuchsia), $-1.4<$[Fe/H]$<-1$ (third panel; green), $-2<$[Fe/H]$<-1.4$ (bottom panel; light-blue).
}
%\vspace{1cm}
\label{fig:phase1}
\end{figure*}
%%%%%%%%%%%%%%%%%%%%%%%%%%%%%%%%%%%%%%
%%% Fig. 3 %%%%%%%%%%%%%%%%%%%%%%%%%%%
\begin{figure*}[hbt!]
\centering
\includegraphics[width=130mm, angle=90]{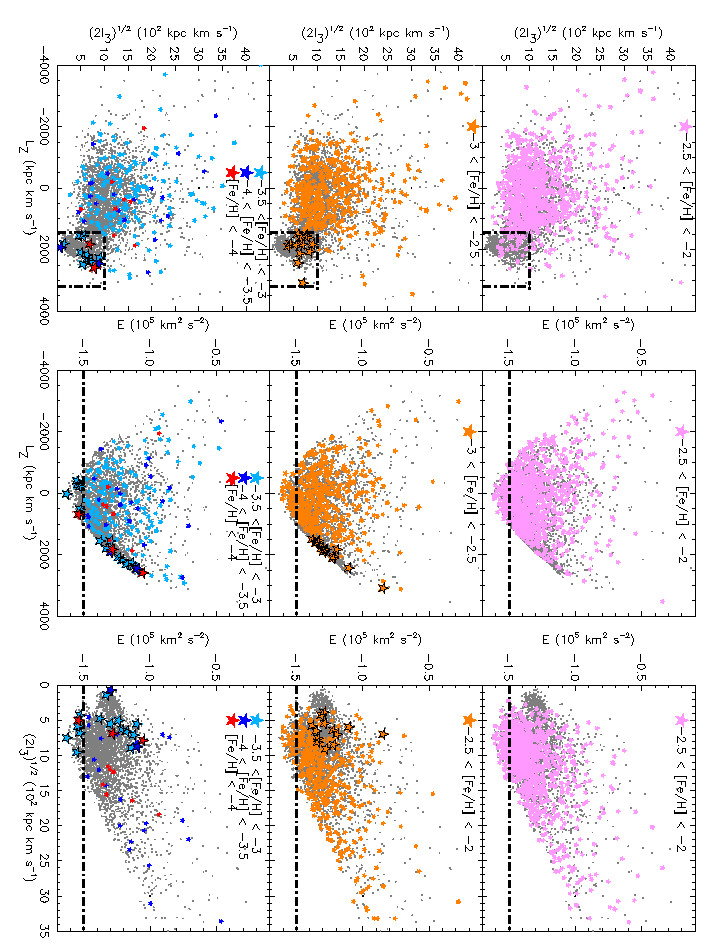}
%\vspace{-4cm}
\caption{
The same as Figure~\ref{fig:phase1} but for the metallicity ranges of
$-2.5<$[Fe/H]$\le-2$ (pink), $-3<$[Fe/H]$\le-2.5$ (orange), $-3.5<$[Fe/H]$\le-3$ (turquoise). The intervals of  $-4<$[Fe/H]$\le-3.5$ and [Fe/H]$\le-4$ are shown in the bottom panels overlapped to the turquoise star symbols, and color-coded with dark-blue and red colors, respectively. In the bottom panels, turquoise, blue, and red filled star symbols with dark edges denote the UMP-EMP stars with disk kinematics (all three panels) and the most-bound VMP-EMP halo stars (middle and right panels). In the bottom-left panel the dotted dashed rectangle shows the ranges of $L_z$ and \I of the  stellar populations in the disks. In the middle and right panels the dashed horizontal lines represents the energy value below which the most bound stars in the Galaxy are located.     
}
\label{fig:phase2}
\end{figure*}
%\vspace{1cm}
%%%%%%%%%%%%%%%%%%%%%%%%%%%%%%%%%%%%%%
\vspace{0.5cm}
\section{Distributions of HES stars in phase-space: dependence on metallicity}\label{Sect:PhaseSpace}

\noindent
In this analysis we make use of the phase-space defined by the three integrals of motion in axisymmetric dynamical models, $L_z$, $E$, and $I_3$. Stellar orbits in this potential are characterized by their distribution in the ($L_z$,$E$,$I_3$) phase-space. In case of the St\"{a}ckel potential, the Hamilton-Jacobi equation separates in ellipsoidal coordinates \citep{dezeeuw1985}, and $I_3$ can be given in analytical form and determined for each star (see CC21, and reference therein). The $I_3$ integral can be considered in the form of $(2I_3)^{1/2}$, which has the dimension of an angular momentum, and correlates well with the maximum orbital angle from the galactic plane (hereafter abbreviated as orbital angle), $\theta_{orb}$. As a  general guide, $(2I_3)^{1/2}$ = 500 \kpckms corresponds to $\theta_{orb} \simeq$ 5$^{\circ}$, $(2I_3)^{1/2}$ = 1000 \kpckms is $\theta_{orb} \simeq$ 15-20$^{\circ}$, and $(2I_3)^{1/2} >$  1000 \kpckms is $\theta_{orb} >$ 20$^{\circ}$. We focus mainly on the global, or “coarse-grained” phase-space distribution of halo stars, where the averaged properties of the halo system, over the phase-space, are close to a dynamically steady state (Binney \& Tremaine 2008).

\noindent
As discussed in \citet{schorck2009}, the sample of metal-poor stars selected in HES with the method described in \citet{christlieb2008} is (intentionally) biased towards extremely metal-poor stars, because the aim of the selection was to efficiently identify candidates for the most metal-poor stars. However, the completeness of the sample is still about 50\,\% at $\mathrm{[Fe/H]}=-2.5$, and it decreases to a few percent at $\mathrm{[Fe/H]}=-2.0$ \citep[][Table 1 and Figure 12]{schorck2009}. Combined with the shape of the MDF of the halo populations, this leads to a considerable number of very metal-poor stars in the HES sample. The present analysis is mainly focused on the VMP halo. However, in the following we perform a phase-space analysis of the full sample, and compare it with some of the main results obtained in CC21.

%%% Fig. 3 %%%%%%%%%%%%%%%%%%%%%%%%%%%
\begin{figure}[t!]
\hspace{-1.1cm}
\centering
\includegraphics[angle=90,width=90mm]{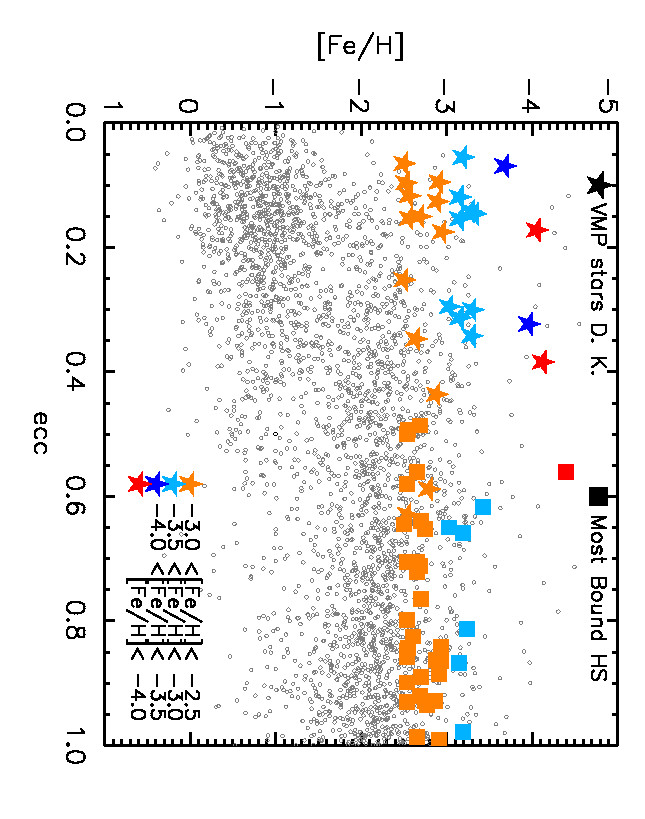}
\caption{Metallicity as a function of the eccentricity for HES stars. The grey dots show the full sample adopted in this analysis, while the color-coded symbols represent different metallicity ranges in the very metal-poor regime. The square symbols denote the most-bound halo stars, while the star symbols represent the VMP stars with disk kinematics (D.K.). 
}
\label{fig: ecc} 
\end{figure}

\noindent
Figures~\ref{fig:phase1} and \ref{fig:phase2} show the distributions of the HES stars in the
phase space defined by the three integrals of motion $(E, L_z, I_3)$, and for nine metallicity intervals.

%[Fe/H]$>-0.6$, $-1<$[Fe/H]$<-0.6$, $-1.4<$[Fe/H]$<-1$, and $-2<$[Fe/H]$<-1.4$ (Figure~\ref{fig:phase1}), and %$-2.5<$[Fe/H]$<-2$, $-3<$[Fe/H]$<-2.5$, $-3.5<$[Fe/H]$<-3$, $-4<$[Fe/H]$<-3.5$, and [Fe/H]$<-4$ %(Figure~\ref{fig:phase2}), for the HES stars. 
% [NC: TOO MANY NUMBERS HERE IN THE TEXT, VERY DIFFICULT TO READ! ALSO, THE INTERVALS CAN BE SEEN IN THE FIGURES.]

\noindent
The grey dots show the full sample, while the color-coded symbols represent sub-samples in the various ranges of
metallicity, as indicated in the legends of each panel. In these figures, the left, middle, and right panels show
the \I \ vs. $L_z$, $E$ vs. $L_z$, and $E$ vs. \I \ diagrams, respectively. Note that, in the metal-poor regime, the bottom of the parabola in the $E$ vs $Lz$ diagram, exhibit a value of $L_z$ = 0 \kpckms, while the bottom of the parabola in the $E$ vs \I diagram, is located at a non-zero value of \I. The lack of stars with $E < -1.4$ and low \I is a general property of metal-poor halo stars, as found in CC21, and it is clearly seen in the HES sample as well.

\noindent
At $\mathrm{[Fe/H]} > -0.6$ (top panels, dark yellow symbols), the sub-sample is dominated by the thin- and thick-disk stellar populations, which possess high $L_z$, low $I_3$, and energy below $-1.1$\,\en. As the metallicity decreases ($-1<\mathrm{[Fe/H]}<-0.6$, 2nd row, magenta symbols), the number of stars with larger values of \I and lower angular momentum increases. This metallicity range is dominated by the overlapping thick-disk and metal-weak thick-disk stellar populations (MWTD; \citealt{carollo2019}, and reference therein), with some halo stars contamination. 
At progressively lower metallicity  ($-1.4<$[Fe/H]$<-1$ and $-2<$[Fe/H]$<-1.4$, two bottom rows, green and cyan symbols), stars exhibit higher values of \I, reaching up to $\sim$ 3000 \kpckms, and lower $L_z$, including many retrograde stars ($L_z <$ 0 \kpckms). In these intermediate metallicity ranges the sample comprises primarily inner halo stars \citep{carollo2007,carollo2010}, that are dominated by the GE debris stars, and whose appearance as elongated feature in the $E$ vs. $L_z$ diagram is visible but less than in the SDSS-DR7 calibration stars sample used in CC21 (see their Figure 3, 3rd and 4th rows of panels). We also notice that in the HES sample the "trail" or elongated distribution of stars with metallicity $-$1.8 $<$ [Fe/H] $< -$1, from $(L_z, (2I_3)^{1/2}) \simeq (1800, 400)$ \kpckms to $(1000, 3500)$ \kpckms, identified in the \I \ vs. $L_z$ diagram by CC21 (and reference therein), is also less evident. The lack of the GE feature and elongated trail is certainly due to the selection in favour of very metal-poor stars in the HES.

\noindent
In the lowest metallicity intervals (Figure 3), the majority of stars have \I $>$ 500 \kpckms and $L_z \lesssim $ 1500 \kpckms, in agreement with what found in CC21, and confirming that, on average, the $L_z$ distribution of the very metal-poor halo is truncated in the prograde high-$L_z$ side, at $L_z \sim$ 1500 \kpckms. In addition, stars with $L_z <$ 1500-2000 \kpckms have always \I $>$ 500 \kpckms (or \tho  $>$ 5-7 deg). The majority of VMP stars possess prograde and retrograde motion (random spin), spanning a large range of values in $L_z$, total energy, $E$, and orbital angles ($I_3$).

%In the middle panel of Figure 3, star symbols color-coded with orange denote stars with %$-$3 $<$ [Fe/H] $\le -$2.5, while in the bottom set of panels, the symbols color-coded %with blue, dark blue and red represent stars with $-$3.5 $<$ [Fe/H] $\le -$3, $-$4 $<$ %[Fe/H] $\le -$3.5, and [Fe/H] $\le -$4. We found N = 10, 9, and 2 stars with $L_z$ %$\gtrapprox$ 1500 \kpckms, and \I $<$ 1000 \kpckms, in these metallicity intervals.
%In fig~\ref{fig:phase2}, bottom ($L_z$, $E$) diagram, we notice, in particular, the two %small clusters of stars at [Fe/H] $< -$4 with similar energy ($E \sim -$1.3 \en), and %located at $L_z \sim$ 1400 \kpckms (3 stars; red filled circle with black color edges), %and 1500$< L_z < $ 2000 \kpckms (3 stars; red filled stars with black color edges). One %star with $-4 < \mathrm{[Fe/H]} < -3.5$ is also associated with this second clump (blue %color). The three stars with $L_z \sim$ 1400 \kpckms appear to be clustered also in the %(\I, $L_z$) diagram, at \I $\sim$ 1400-1500 \kpckms and they have halo kinematics, %given their higher \I value. Hereafter, we refer to this group of stars as simply "UMP %halo clump". The group of stars with larger vertical angular momentum exhibit %increasing \I along an elongated feature visible in both the \I vs $Lz$, and $E$ vs \I %diagrams. We refer to this group of stars as UMP-EMP halo clump with disk kinematics.

%%%%%%%%%%%%%%%%%%%%%%%%%%%%%%%%%%%%%%
\noindent
However, a few stars, possess $L_z$ $\gtrapprox$ 1500 \kpckms, and \I $<$ 1000 \kpckms, which are typical values for stars in the disk populations.  \citet{sestito2020} report the existence of very metal poor stars ([Fe/H] $< -$2.5) with disk kinematics in the Pristine survey \citep{starkenburg2017}. They also found an asymmetry between prograde and retrograde disk-like stars in this metallicity regime, with the prograde region much more populated than the retrograde counterpart.  

\begin{figure}[t!]
\hspace{-1.1cm}
%\centering
%\includegraphics[scale= 3.5]{figure5.jpg} % The file of Figure 3 is called "figure5.pdf"?
\includegraphics[width=95mm]{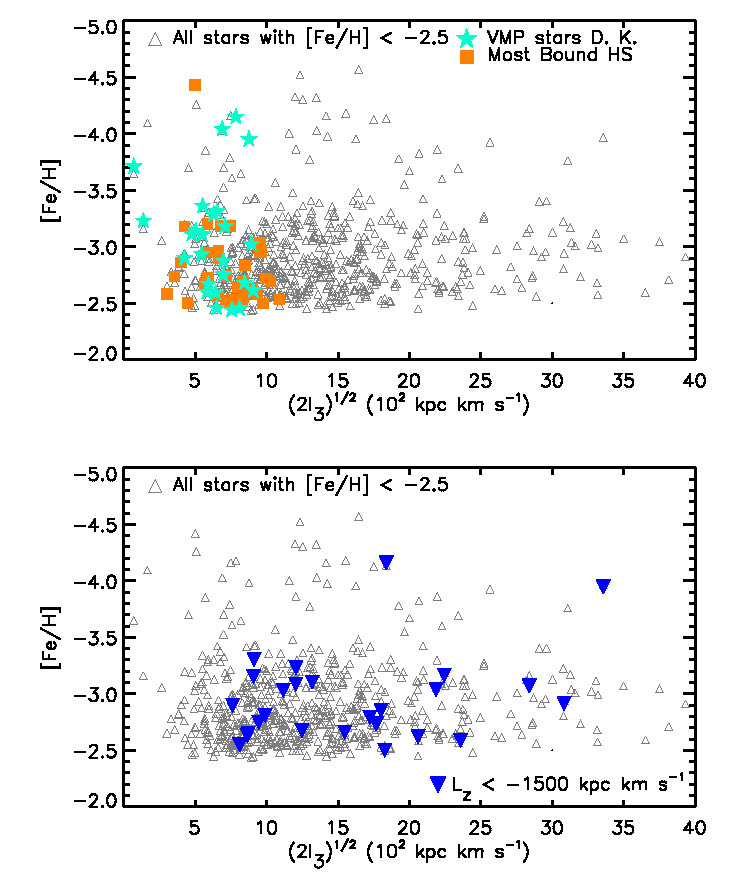}
\caption{Top-panel: Metallicity as a function of \I for stars with [Fe/H] $< -$2.5 (gray unfilled upward triangles). Orange square symbols represent the most-bound halo stars, while the turquoise star symbols denote the VMP stars with disk kinematics (D.K). Bottom panel: Metallicity as a function of \I for stars with [Fe/H] $< -$2.5 (upward gray triangles) overlapped with stars with highly retrograde motion and ($L_z < -$1500 \kpckms; dark blue downward triangles).
}
\label{fig:ecc1} % Don't use blanks in label commands! 
\end{figure}

\noindent
Figure~\ref{fig:phase2} shows that the HES sample contains such stars and, to get further insight, we plot, in all the panels of this figure a dotted-dashed rectangle (left column of panels), which represents the region in the \I vs $L_z$ diagram populated by stars with disk kinematics (thin- and thick-disk), and corresponding to $L_z >$ 1500 \kpckms and \I $<$ 1000 \kpckms. Indeed, CC21 have shown that, in the prograde region of the \I vs $L_z$ diagram, halo stars can be separated from the disk stellar components by adopting, 0 $< L_z <$ 1500 \kpckms and \I $>$ 1000 \kpckms. Therefore, the majority of stars inside the rectangle and with metallicity [Fe/H] $> -$1, belong to the disk populations (thin- and thick-disk). Examination of Figure 3 reveals that some of the HES stars with [Fe/H] $< -$2 possess disk kinematics. In particular, we found 33, 13, 9, 2 and 2 stars, in the metallicity ranges of $-2.5<$[Fe/H]$\le -2$, $-3<$[Fe/H]$\le -2.5$, $-3.5<$[Fe/H]$\le -3$, $-4<$[Fe/H]$\le -3.5$, and [Fe/H]$\le-4$, respectively. The associated \I values are in the range of 100 $<$ \I $<$ 900 \kpckms, ($1^\circ \lesssim \theta_{\rm orb} \lesssim 12^\circ$). These orbital angles correspond to values of the maximum distance from the Galactic plane of 0.2 kpc $\lesssim$ $z_{\rm max} \lesssim$ 3 kpc. 

\noindent
Note that the MWTD possesses an average vertical angular momentum of $\langle L_z \rangle$ $\sim$ 1200 \kpckms, with a dispersion of $\sigma_{\rm Lz}$ $\sim$ 500 \kpckms. Moreover, as shown in CC21, MWTD stars exhibit values of \I higher than the thick-disk, \I $\sim$ 850 \kpckms for the MWTD, compared to the average value for the thick disk of \I $\sim$ 500 \kpckms, which implies that the MWTD has systematically larger orbital angles than those possessed by the thick disk ($\theta_{orb}$ $\sim$ 10 deg for the MWTD and $\theta_{orb}$ $\sim$ 7 deg for the thick disk). Inside the dotted-dashed area, at  $L_z >1500$\,\kpckms, we do not expect to find a significant fraction of MWTD stars, because they lag behind this $L_z$ limit and possess orbital angles larger than the canonical disk populations.

\noindent
In the rest of this analysis, we will consider only stars with metallicity [Fe/H] $< -$2.5, instead of [Fe/H] $< -$2.0, which is the typical upper limit for VMP stars. This choice takes into account the uncertainty on the metallicity determination ($\sim$ 0.3), and the higher completeness of the sample in this metallicity range.

\noindent
We found that the number of stars with [Fe/H] $\lesssim -$2.5 and disk kinematics is more than three times its retrograde counterpart (N$_{\rm Prograde}$ = 26, N$_{\rm Retrograde}$ = 8), in agreement with \citet{sestito2020}.\footnote{At [Fe/H] $\lesssim -$2.0 these numbers are N$_{\rm Prograde}$ = 61, N$_{\rm Retrograde}$ = 23} 
%At [Fe/H] $< -$2 these numbers become N$_{\rm Prograde}$ = 144, N$_{\rm %Retrograde}$ = 77. 
As discussed in CC21 (and reference therein), a fraction of these stars were likely formed in-situ from infalling early and pristine cooled gas settled into an equatorial plane of a progenitor dark halo, where an angular momentum was present. Candidates in-situ halo stars can be also defined as those being most tightly bound to the Milky Way gravitational potential and possessing the lowest binding energy. These are stars that ended up in the bottom of the main progenitor halo during the early stages of its formation. A fraction of such stars formed from cooled gas, or from merged/accreted cold gas supplied by other halos. and they are the most tightly bound to the gravitational potential of the progenitor halo after dissipative cooling (CC21). Therefore, candidate in-situ halo stars can be defined as those possessing the lowest binding energy.  
Hence, both groups of stars can be considered dynamically informative of the assembly history of the Galaxy.

\noindent
In Figure~\ref{fig:phase2}, the horizontal line in the middle and right panels shows the locus of stars with the lowest binding energies, $E < -$1.5 \en, and $Lz \sim$ 0 \kpckms. In CC21 was shown that stars with such low energy populate the third integral of motion in the range 400 $\lesssim$ \I $\lesssim$ 1200 \kpckms, and stars with the lowest energy have \I $\sim$ 700-900 \kpckms. The orbital angle associated to these values of \I are , $5^\circ \lesssim \theta_{\rm orb} \lesssim 20^\circ-30^\circ$, and $\theta_{\rm orb} \simeq 10-12^\circ$ for the lowest $E$. When expressed in terms of the maximum distance from the Galactic plane, $z_{\rm max}$, these stars have $z_{\rm max} < 4$--$5$\,kpc. 

\noindent
In the HES sample we found 31, 5, and 1, most bound halo stars, in the metallicity intervals $-$3.0 $<$ [Fe/H] $< -$2.5, $-$3.5 $<$ [Fe/H] $< -$3.0, and [Fe/H] $< -$4.0, respectively, while there are no such stars in the range $-$4.0 $<$ [Fe/H] $< -$3.5.  

%Hereafter, we name candidates in-situ type-1 halo stars, those with the lowest binding %energy, while candidates in-situ type-2 halo stars are those with $\mathrm{[Fe/H]} < -2.5$ %and disk kinematics, keeping in mind that a significant fraction of the type-2 stars could %have accreted origin.

%The clump with disk kinematics and [Fe/H] $< -$3.5 is %associated to the in-situ halo type-2.  
\noindent
Figure~\ref{fig: ecc} (top panel), shows the metallicity as a function of the eccentricity for the full HES sample (grey dots). Star and square symbols represent VMP stars with disk kinematics, and the most bound halo stars, respectively, for different ranges of metallicity, starting from $\mathrm{[Fe/H]} \sim -2.5$, and below. The diagram shows that the majority of the most-bound halo stars exhibit high orbital eccentricities in the range of $0.5 < \mathrm{ecc} < 1$, while highly prograde VMP stars have low orbital eccentricities, $\mathrm{ecc} < 0.4$, as expected for stellar populations with disk kinematics. The rest of the stars with metallicity $\mathrm{[Fe/H]} < -2.5$ possess a uniform distribution of eccentricities ranging from $\mathrm{ecc} = 0$ to 1.

%Stellar members of the EMP-UMP halo cluster with disk kinematics have %eccentricity in the range $0.08 < \mathrm{ecc} < 0.35$, and are evidenced with %large unfilled star symbols (blue and red). The red filled circles denote the %location of the UMP halo clump, which possesses low eccentricity, in the range %of $0.1 < \mathrm{ecc} < 0.25$.

%\\ When you want to start a new paragraph, just insert a blank line! 

%An important property of metal-poor halo stars is the fact that they %always have \I $\gtrapprox$ 500 \kpckms (\tho  $>$ 5-7 deg). It is %interesting to note that some VMP stars with disk kinematics do not adhere %to this property as can be clearly seen in the top panel of %Figure~\ref{fig: ecc1}, which represents the metallicity as a function of %\I, for stars with [Fe/H] $< -$2.5 (gray unfilled upward triangles).

\noindent
Figure~\ref{fig:ecc1} shows the metallicity as a function of \I, for stars with [Fe/H] $< -$2.5 (gray unfilled upward triangles). Turquoise and orange filled squares denote VMP stars with disk kinematics, and the most-bound halo stars, respectively. The majority of such stars possess 400 $<$ \I $<$ 1000 \kpckms (5-7 deg $<$ \tho  $<$ 20-30 deg), with the exception of two VMP stars with disk kinematics, which can have \I values as low as $\sim$ 100 \kpckms, (orbital angles 1-2 deg).

\noindent
Figure~\ref{fig:ecc1} also shows that the majority of stars in the very metal-poor halo exhibits values of \I in the range $\sim$ 500 \kpckms $<$ \I $<$ $\sim$ 2000 \kpckms (80\% of the sample), corresponding to orbital angles in the range of 5-7 deg $<$ \tho $<$ 30-40 deg . The remaining stars have \I $\gtrsim$ 2000 \kpckms that is, orbital angles of \tho $>$ 40-50 deg (20\%).

\noindent
The bottom-panel of Figure~\ref{fig:ecc1} shows [Fe/H] vs. \I for stars with [Fe/H] $< -$2.5 (gray unfilled upward triangles) overlapped with stars possessing highly retrograde motion ($L_z < -$1500 \kpckms; dark blue downward triangles) including those with \I $<$ 1000 \kpckms and representing the retrograde counterpart of stars with disk kinematics, in the same range of metallicity. Inspection of this panel reveals that all the highly retrograde VMP stars have always \I $>$ 8-10 (10$^2$ \kpckms), or \tho $>$ 20-30 deg.

\noindent
Note that our full sample contains only 10 stars at $\mathrm{[Fe/H]}\le -4$. The low-metallicity tail of the MDF decreases roughly by a factor of 10 for every dex in [Fe/H] \citep[e.g.][]{schorck2009}. The full sample analysed here contains 228 stars at $\mathrm{[Fe/H]}\le -3$ (see Tab.~\ref{Tab:SampleStatistics}), hence $\sim$ $23\pm 5$ stars at $\mathrm{[Fe/H]}< -4$ would be expected by applying aforementioned scaling, which is consistent with the actual number of stars at this metallicity in our sample.

%There are also only very few stars with \I $>$ 2000-2500 \kpckms. This is probably due to the fact that the faintest stars in our sample have $B\approx 18$, corresponding to $G\approx 17$ so that for cool giants, only distances up to ???\,kpc [DANIELA, PLEASE INSERT NUMBER] can be reached. 

%%% Fig. 6 %%%%%%%%%%%%%%%%%%%%%%%%%%%
%% [NC: NO, THIS IS FIGURE 7!]

\begin{figure*}[t!]
\hspace{2cm}
%\vspace{0.5cm}
%\centering
%\includegraphics[scale= 3.5]{figure5.pdf} 
\includegraphics[angle=0,width=140mm]{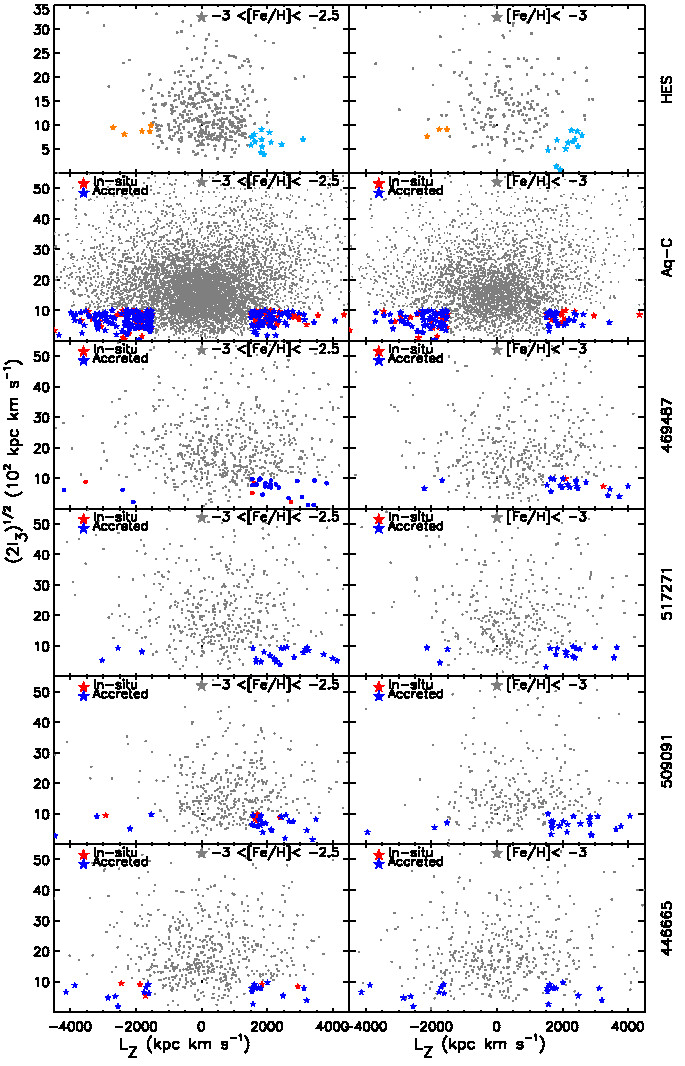} 
\caption{($L_z$,\I) diagram for VMP HES stars (top panel) in two ranges of metallicity below [Fe/H] = $-$2.5. Cyan symbols denote stars with disk dynamics, while orange symbols show the retrograde counterpart. Second panel: ($L_z$,\I) diagram for VMP stars in Aq-C simulated halo. Panels 3-6: ($L_z$,\I) diagram for TNG50 MW's analog galaxies in the same metallicity intervals. In all the panels gray star symbols show the subsamples in each metallicity range, while blue and red star symbols denote stellar particles with accreted and in-situ origin, respectively. 
}
\label{fig:simulations}
\end{figure*}

%\newpage
\section{Implication for the formation of the early Galaxy}

\noindent
The chemo-dynamical analysis of the HES full sample shows that the VMP stellar halo is made of both stars with prograde and retrograde motion, spanning a large range of values in $L_z$, total energy, $E$, and orbital angles ($I_3$). In this metallicity range, the majority of stars are characterized by $L_z <$ 1500 \kpckms, and \I $>$ 500 \kpckms, with a discontinuous distribution at $L_z \simeq$ 1500 \kpckms, and \I $\simeq$ 500 \kpckms. 
%According to \citet{carollo2007} and CC21, these properties are intrinsic of %and suggest that this layer of the halo is accreted dominated and
%likely formed from a chaotic, dissipationless random merging and
%accretion of low-mass satellites whose candidate surviving
%counterparts could be the ultra-faint dwarf spheroidal galaxies
%(UFD).\\
%Numerical simulation of stellar halo predict  the formation of stellar halos %mainly by accretion of satellites galaxies\citep[e.g.][]{zolotov2009,font2011, %tissera2012}\\
\noindent
The HES sample has also revealed the presence of VMP stars with disk kinematics ($L_z >$ 1500 \kpckms and \I $<$ 1000 \kpckms) with metallicity as low as [Fe/H] $< -$4. 

\noindent
In order to get insight on the main properties of
these peculiar stars, as well as the most bound VMP halo stars, we have analyzed simulated galaxies with properties similar to the MW,  obtained from two different sets of simulations: the Aquarius project \citep{scannapieco2009}, and the TNG50 simulations \citep{pillepich2019a,nelson2019b}. We explore the presence of these dynamically informative stars in such simulated galaxies aiming at understand their origin. The comparison of the observational findings with simulated halos obtained from numerical simulations that adopt different sub-grid physics, and codes, is an important step to probe some level of robustness of the results, since the low-metallicity regime can be elusive and difficult to describe, because of the limitations of numerical resolutions.

%In the following subsections, we describe the methodology used to select %stellar particles in simulated galaxies, which are consistent with the %observational criteria adopted before, and describe the main characteristics %of
%the analysed halos in Aq-C and the Illustris-TNG50 galaxies, and compare with %the observations.

%and  with similar properties and their origin.
%, however, a detailed
%analysis of their origin  is beyond the scope of this work.
%We are particularly interested in halo stars with disk kinematics and the most bound stars.
%, a fraction of which can be considered candidates in-situ halo stars, according to CC21. 

\subsection{Aquarius C MW analog}

\noindent
Briefly, the Aquarius Project consists of 8 MW mass-sized halos
\citep{scannapieco2009}. The initial conditions were selected from the dark matter only simulation consistent with a Lambda Cold Dark Matter ($\Lambda$CDM) scenario, $\Omega_{\Lambda}= 0.75$,$\Omega_{m} = 0.25,\Omega_{b} = 0.04, \sigma_8 = 0.9$ and H$ _0=73$ km s$^{-1}$ Mpc$^{-1}$ (Aq-C has been re-scaled to match the TNG50 cosmology). The mass of the dark matter and initial gas particles are  $10^{6}$~${\rm M_\odot }$~$h^{-1}$ and 
$2\times$10$^{5}$~${\rm M_\odot}$~$h^{-1}$, respectively. The simulated  halos were selected to have virial masses in the range $7-16 \times 10^{11}$M$\sun$ and not have had a major merger since $z < 2$. 

\noindent
The stellar halos of the Aquarius galaxies have been shown to be formed mainly by the accretion of satellite with a variety of stellar masses, which brought stars and gas with them. In the inner region, a contribution of an in-situ component was also found, the majority of which was produced by the dynamical heating of the disc components \citep{tissera2012,tissera2013}. The metallicity profiles of the halos as a function of galactocentric distance is determined by  the mass function of the accreted satellite and their orbital parameters \citep{amorisco2017,fernandez-alvar2019}. The outer regions of the halos show a more significant contribution from small satellites (M$_{\ast} < 10^9$ M$_{\sun}$) compared to the inner regions which are built up by larger inputs from a few massive satellites \citep{tissera2014,tissera2017}. The fraction of VMP stars is found to increase with increasing galactocentric distance, with 15--20\ \% of the stellar populations with [Fe/H]$ < -2$ at around 10 kpc from the galactic centre. These stars are formed both in-situ and accreted, with the later being older and more $\alpha$-enriched \citep{tissera2017}. 

\noindent
 In this work, we focus on Aquarius C (hereafter, Aq-C) that has been reported to  better reproduce  properties of the MW galaxy, such as the chemo-dynamical properties \citep{tissera2013,tissera2017}, age distributions \citep{carollo2019,whitten2019} and the [$\alpha$/Fe] radial distributions in the inner halo region \citep{fernandez-alvar2018}. Additionally, \citet{Whitten2021} reported that the VMP in the inner region of Aq-C are mostly high $E$ stars that happened to be in the central regions due to their large eccentric orbits.

\noindent
 To analyse the origin of stellar populations, we use the classification of in-situ, endo-debris and accreted stars adopted by \citet{tissera2013}. According to these authors, in-situ stars are those formed in the main progenitor galaxy, endo-debris stars formed inside the virial radius of a galaxy from gas brought in by satellites, and accreted stars are those that formed in satellites before falling within the virial radius of the progenitor system. \citet{tissera2013} followed back all star particles to their site of formation, identifying the system where they formed and the time at which these systems entered the virial radius of the progenitor galaxy (i.e. the progenitors are defined as the most massive system in the merger tree). For simplicity, in this paper, endo-debris and accreted stars will be put together under the label of accreted stars. 
 \vspace{0.5cm}

%\newpage
\subsection{TNG50 MW Analogs}

\noindent
IllustrisTNG is a series of large, cosmological magnetohydrodynamical simulations of galaxy formation based on AREPO \citep{springel2010}. The simulations provide three physical box sizes, and cubic volumes of 50 Mpc (TNG50), 100 Mpc (TNG100), and 300\,Mpc (TNG300) side length. The simulations are consistent with $\Lambda$CDM cosmology with the following parameters: $\Omega_{\Lambda,0}$ = 0.6911, $\Omega_{m,0}$ = 0.3089, $\Omega_{b,0}$ = 0.0486, $\sigma_{8}$ = 0.8159, $n_{s}$ = 0.9667, $h$ = 0.6774, consistent with constraints provided by \citet{planck2016} (see also, \citealt{pillepich2018}, \citealt{nelson2018}, \citealt{naiman2018}, \citealt{marinacci2018}, \citealt{springel2018}).

%TNG includes a comprehensive model for
%galaxy formation physics, which follows the formation and evolution of galaxies across cosmic time 
\noindent
TNG50 is the high resolution simulation of the TNG series with $m_{\rm DM} = 4.5 \cdot 10^{5}$ M$_{\odot}$, and $m_{\rm baryon}
= 8.5 \cdot 10^{4}$ M$_{\odot}$. More details can be found in  \citealt{pillepich2019a},  \citealt{nelson2019b},  \citealt{engler2021}, and  \citealt{pillepich2021}.\\
 MW analogs  are selected on the basis of their stellar mass, shape, and kinematic properties. The total stellar mass within an aperture of 30 kpc is required to be in the range M$_{\star} = 5 \cdot 10^{10} - 9 \cdot 10^{10}$ M$_{\odot}$\footnote{The MW total baryon mass is $\sim$ 6.7$\cdot$10$^{10}$M$_{\odot}$, \citealt{flynn2016}}, and galaxies must have a disky shape determined by having a minor-to-major axis ratio of their 3D stellar mass distribution of c/a $<$ 0.45 (measured between one and two times the stellar half-mass radius; \citealt{engler2021}, and references therein). Simulated galaxies that satisfy the c/a properties are also visually inspected through the synthetic three-band images in edge-on and face-on projections, and they have to appear of disky shape and exhibits spiral arms. MW analogs are also selected such that no other galaxy more massive than 10$^{10.5}$ M$_{\odot}$ is within a 500\,kpc distance, and the total mass of the halo host is not typical of a massive group and cluster, i.e. M200(host) $<$ 10$^{13}$ M$_{\odot}$.

\noindent
Additionally, the distribution of stellar particles in the ($L_z$,$L_p$) plane should resemble that of the MW, with highly rotating disk(s) components and, on average, a non-rotating halo. 
We selected four galaxies that satisfy all of the above criteria, with identification numbers, IDs, 469487, 517271, 509091 and 446665. 

%\subsubsection{Classification of in-situ and accreted stars in the Illustris-TNG50 MW's analogs}
\noindent
We used the public catalog\footnote{https://www.tng-project.org/data/} \citep{nelson2019a} to access the halos, the central and satellites galaxies, as well as their merger trees. In order to make a similar classification according to the site of formation as used for Aq-C, stellar particles are followed back in time to find the systems where they were born.
If this occurs in satellite galaxies that later merged with the central one, the stars are classified as accreted.
However, if stars are born from gas present within the virial radius of the central galaxy, we need to identify whether or not such gas is attached to an infalling satellite that could not be disentangled by the SUBFIND algorithm\footnote{We note that halos and galaxies in both simulations have been identified by applying the Friends-of-Friend and the SUBFIND algorithms (FoF; \citealt{davis1985} and SUBFIND; \citealt{springel2001a, dolag2009})}. In case the gas where the stellar particles are born is attached to the satellite galaxy, such particles are classified as accreted, otherwise they are classified as in-situ.

\noindent
To perform this classification, we applied a new algorithm that allows us to isolate stellar particles that belong to a physically separated structure and located close to the main galaxy, with a confidence level of $\sim 2 \sigma$. This stellar particles could not be individualized by the SUBFIND algorithm (Sillero et al. in preparation). To do this, in each subhalo, we use all the selected star particles and analyze their potential energy distribution as a function of the distance from the barycenter of the main galaxy. The barycenter is obtained iteratively  and coincides, within a small dispersion, with the location of the  most bound particle in a given subhalo. Then, by examining the dominant trend in the aforementioned potential energy distribution, we can identify particle agglomerations associated with the presence of different compact structures and confine them spatially. This method makes possible the identification of close passages, or advance merger phases.

\noindent
This procedure is not applied to Aq-C because \citet{tissera2013} identified the stellar particles formed in infalling satellite galaxies, and their parent gas particles, by following them back in time along the merger trees. These particles are classified as endo-debris by Tissera et al. In this paper, the endo-debris star particles are considered as accreted  ones in order  to be consistent with the TNG50 classification explained above. Hence, although the methods are different their results are comparable.

\subsection{Defining the simulated counterparts of the observed VMP samples}

\noindent
 We apply the same analysis implemented for the observed VMP stars in the HES sample to the simulated data at $z$ = 0 . For all the stellar particles in the halos of simulated galaxies, and for both simulations, we determined $L_z$ and \I. To do this, we work with the rotated galaxies so that the galactic plane coincides with the x-y plane, and the total angular momentum of the disk component is aligned with the z-axis. The \I parameter is evaluated in a MW's potential of a St\"{a}ckel shape, as for
the observed data, and knowing that this type of potential is a good approximation for many galactic mass distributions. This also ensure that both the simulated and observational parameters are estimated under similar hypothesis.

\noindent
 After the transformation from Cartesian to Galactocentric cylindrical coordinates, we selected stellar particles in an annulus of 6\,kpc $<$ ${\rm R}$ $<$ 20\,kpc, where {\rm R} is the projected galactocentric distance (assuming the Sun position at ${\rm R}$ = 8.5 kpc). This range is more extended than the one adopted for the MW, and takes into account that the analysed galaxies are only MW's analogs, and hence, the solar neighbourhood might not be at the same distance from the galactic centre as in the MW.
 
 %The maximum height above the
 %plane considered in this work is Z $\sim$ 50 kpc, which mimics well %the region covered by observations.\\ 
\noindent
For each stellar particle, we estimate the orbital
parameters, r$_{\rm apo}$, r$_{\rm peri}$, eccentricities defined as in Sect. 2, and the integrals of motion $L_z$, \I, $E$. Metallicity ([Fe/H]), $\alpha$-elements abundance ([O/Fe] or [Mg/Fe]), and ages are also known for each stellar particle in both simulations.

We clarify that, after stellar particles are selected and classified by using the observational criteria, they are followed back in time until their birth location. At each available redshift of the simulations, the progenitor galaxies are rotated so that the z-axis is aligned with the direction of the angular momentum of the stellar component. The circularity parameter $\epsilon= L_z/L$ is estimated for all stellar particles and gas particles so that we can identify if the stellar particles of interest were born from gas rotational supported at that time. This has been done when there were around 500 particles in the progenitor to robustly estimate the angular momentum (e.g. \citealt{rodriguez2022}). We also search for those born in a separate galaxy that later will be accreted by the progenitor, following them as they spiral into the potential-well of the progenitor system as described in the previous subsections.

%In both Aq-C and TNG50 galaxies, the stellar parameters provided for each galaxy are: galactocentric cartesian coordinates, ($x$,$y$,$z$), metallicity ([Fe/H]), $\alpha$-elements ([Mg/Fe]), age, galactocentric velocity components, ($v_x$,$v_y$,$v_z$), and particle IDs.
\vspace{0.5cm}
\subsection{VMP stars with disk kinematics}

\noindent
To search for simulated stellar populations that have similar chemo-dynamical properties as the VMP stars with disk kinematics (and their retrograde counterpart), identified in HES, 
%of halo stars in MW disks (and their retrograde counterpart), 
we performed the same selection used for the HES sample in the phase space defined by ($L_z$, \I).  Figure \ref{fig:simulations} shows the comparison between VMP stars in the HES sample, Aq-C, and the four TNG50 MW' analogs.
%In the simulated halos, we found that the number of star particles that satisfy the
%criteria to be classified as VMP is quite low and hence, the following %analysis has
%to be taken as indicative. 
\noindent
The upper panel of Fig.~\ref{fig:simulations} shows \I as a function of $L_z$ for the HES VMP sample divided in two ranges of metallicity: $-3 < \rm{[Fe/H]}<-2.5$ and $\rm{[Fe/H]} <-3$ (left and right top panel; gray star symbols). Stars with disk kinematics and its retrograde counterpart are color-coded in cyan and yellow, respectively. We include this plot to facilitate the comparison with observations.

\noindent
The second row of panels in Figure~\ref{fig:simulations} shows similar diagrams for the simulated Aq-C VMP halo while the rest of them display similar information for the four selected TNG50 MW analogs. Stellar particles with disk kinematics are color-coded according to their origin: red are formed in-situ and blue are accreted. The diagrams also show the retrograde counterpart where the same color-codes are adopted.

%In the simulated halos, stellar particles with disk kinematics are selected by adopting the criteria defined in CC21.
\noindent
In both simulated halos and observations, we identify two subsamples: VMP stars with disk kinematics and its retrograde counterpart. To quantify the trends, we estimate the fractions of HES halo stars, and the stellar mass fractions  in Aq-C and TNG50 MW's analogs in  the mentioned subsamples.

%Table\footnote{We note that in the Table 2 fractions are shown as percentages.} 
\noindent
Table \ref{Tab:Simulations1} summarizes such fractions showed as percentages for the two subsamples, in the two  defined metallicity ranges, $-$3 $<$ [Fe/H] $\leq -$2.5, and [Fe/H] $\leq -$3. The HES fraction is determined by dividing the number of VMP stars with disk kinematics by the total number of stars in each metallicity range. In case of Aq-C and TNG50 MW' analogs, the stellar mass fraction is defined as M$_{\rm F}$ = $\sum_{i} m_{i}$ / $\sum_{k} m_{k}$, where $i$ runs within each subsample (i.e. disk kinematics and retrograde counterpart), and $k$ runs over the total sample in each metallicity range.

\noindent
Inspection of Table \ref{Tab:Simulations1} reveals that in the HES sample, the fraction of prograde stars is larger than its retrograde counterpart, in particular, in the lower metallicity range. A similar trend is confirmed for the simulated galaxies with the
exception of Aq-C, and the lower metallicity range of the 446665 TNG50 galaxy, which exhibit a larger fraction of stellar mass in the retrograde counterpart. 

\noindent
In Aq-C the fraction of VMP stellar mass with disk kinematics is
lower than its retrograde counterpart in both ranges of metallicity,
%($N_P$ = 50, $N_R$ = 75 at $-$3 $<$ [Fe/H] $< -$2.5, and $N_P$ = 38,$N_R$ = 81 at [Fe/H] $< -$3), 
in contrast with the trend observed in the HES sample. This could be due to the fact that Aq-C has experienced a massive retrograde merger, which could have left a more important contribution of low metallicity stars in the halo with retrograde motion \citep{fernandez-alvar2019}.

\noindent
In case of TNG50 MW halos, in the higher metallicity range (left column of panels in Fig.~\ref{fig:simulations}), the fraction of stellar mass with prograde motion tends to be larger than the retrograde counterpart (from five to eight times), with the exception of galaxy 446665, where the distribution is $\sim$ 50\%-50\%. This tendency is also confirmed in the lower metallicity range (right column of panels in Fig.~\ref{fig:simulations}), and it is in agreement with the trend observed in the top panels for the HES sample, as well as in the sample of \citet{sestito2020}. 

\noindent
It is important to note that, for both Aq-C and TNG50 MW analogs, the number of stellar particles in both low-metallicity subsamples is not very high, in particular, for the TNG50 galaxies ({\it N} = 300-700). Such numbers depend on both the history of assembly of each halos and the numerical resolution. When the low metallicity subsamples are split into the different categories described in Tables 2, 3, 4, and 5 the number of particles decreases further.
In order to obtain a more robust estimation of the signals we determined the bootstrap errors of the stellar mass fraction for each of the reported subsamples ($>$ 100 random samples for each metallicity interval), in each of the categories described in the tables. When the estimated mass fraction are within three bootstrap $\sigma$, the correspondent signal, or result, might be consistent with numerical noise. These very few cases are shown in boldface in the tables. Such results suggest that the trends reported in the tables are statistically reliable, although they should be taken as indicative. Nevertheless, the similarity of the global trends in these diverse MW analogs strengthens our analysis, in particular, considering that each galaxy has a different assembly history, and the simulated galaxies are products of different numerical codes and subgrid algorithms.

%In the Tables, the estimated mass fractions which are within three bootstrap $\sigma$ are shown in boldface.
%\textcolor{blue}{Hence, for these cases the signal might be consistent with numerical noise.}

%{\bf In the range of metallicity of $-$3 $<$ [Fe/H] $\leq -$2.5 we found $N_{\rm D.K.}$ = 50, 25, 22, 28, 13 and $N_{\rm Ret.  C.}$ = 75, 4, 3, 5, 12, while at [Fe/H] $\leq -$3 the numbers are, $N_{\rm D.K.}$ = 38, 22, 15, 21, 8 and $N_{\rm Ret.  C.}$ = 81, 2, 3, 3, 9 for Aq-C, TNG50-469487, TNG50-517271, TNG50-509091, TNG50-446665, respectively (D.K. denote Disk Kinematics and Ret. C. stands for Retrograde Counterpart). $-->$ \DCb{Patricia here we have to comment on how these low numbers affect the results.}}
%In general, the trends are reliable statistically, although should be taken  as indicative. 
%\textcolor{green}{\bf However, it strengthens our analysis that global trends are similar,}
%Nevertheless, it is reassuring, that the global trends are similar, 
%considering that each galaxy has a different assembly history, and the simulated galaxies are %products of different numerical codes and subgrid algorithms.
%}

%\DCb{Why do we say this is most of the mass f. are noise only} \\
%, and taking into account codes (including subgrid physcs) adopted to run the %analysed simulations.

\begin{deluxetable*}{llcccccc}
\tabletypesize{\footnotesize}
\singlespace
\tablecolumns{12}
\tabcolsep=0.07cm
\tablewidth{0pt}
\label{Tab:Simulations1}
\tablecaption{Fractions of VMP stars (HES) and stellar mass fractions of simulated galaxies (Aq-C and TNG50) for sub-samples with disk kinematics and their retrograde counterpart.}
\tablehead{
\colhead{Metallicity} &
\colhead{Rotation} &
\colhead{HES} &
\colhead{Aq-C} &
\colhead{469487} &
\colhead{517271} &
\colhead{509091} &
\colhead{446665} \\
\colhead{} &
\colhead{} &
\colhead{F (\%)} &
\colhead{M$_{\rm F}$ (\%)} &
\colhead{M$_{\rm F}$ (\%)} &
\colhead{M$_{\rm F}$ (\%)} &
\colhead{M$_{\rm F}$ (\%)} &
\colhead{M$_{\rm F}$ (\%)} \\
}
\startdata 
$-$3 $<$ ${\rm [Fe/H]}$ $\leq -$2.5    & Disk kinematics         & 2.8 & 1.2 & 3.5 & 4.3 & 6.1 & 1.9  \\ 
                                       & Retrograde counterpart  & 1.1 & 1.8 & 0.5 & 0.6 & 0.9 & 1.7  \\
\hspace{0.7cm} ${\rm [Fe/H]} \leq -$3  & Disk kinematics         & 7.0 & 1.0 & 4.3 & 4.0 & 7.3 & 1.8  \\
                                       & Retrograde counterpart  & 1.4 & 2.1 & 0.2 & 1.0 & 0.9 & 2.7\\
\hline
\hline
\enddata
%\tablecomments{}
\end{deluxetable*}

\begin{deluxetable*}{lllcccccccccc}
\tabletypesize{\footnotesize}
\singlespace
\tablecolumns{13}
\tabcolsep=0.11cm
\tablewidth{0pt}
\label{Tab:Simulations2}
\tablecaption{Stellar mass percentages of in-situ and accreted VMP components with disk kinematics and their retrograde counterpart. The corresponding median ages are also shown. Boldface numbers denote mass fractions within three bootstrap $\sigma$.}
\tablehead{
\colhead{Metallicity} &
\colhead{Rotation} &
\colhead{Origin} &
\colhead{Aq-C} &
\colhead{Age} &
\colhead{469487} &
\colhead{Age} &
\colhead{517271} &
\colhead{Age} &
\colhead{509091} &
\colhead{Age} &
\colhead{446665} &
\colhead{Age}\\
\colhead{} &
\colhead{} &
\colhead{} &
\colhead{M$_{\rm F}$ (\%)} & 
\colhead{Gyr} &
\colhead{M$_{\rm F}$ (\%)} &
\colhead{Gyr} &
\colhead{M$_{\rm F}$ (\%)} &
\colhead{Gyr} &
\colhead{M$_{\rm F}$ (\%)} &
\colhead{Gyr} &
\colhead{M$_{\rm F}$ (\%)} &
\colhead{Gyr}
}
\startdata 
                                     & Disk kinematics & In-situ  & 0.4 &  12.2  & 0.9  & 12.9  & $-$  & $-$  & 0.8 & 12.7 & {\bf0.3 } & 12.5 \\
$-$3 $<$ ${\rm [Fe/H]}$ $\leq -$2.5  & Disk kinematics & Accreted & 0.8 & 12.7   & 2.6  & 12.9  & 4.3  & 12.8 & 5.3 & 12.8 & 1.6  & 12.5  \\
                         & Retrograde Counterpart & In-situ  & 0.5 &  13.5  & {\bf 0.3}  & 12.5  & $-$  & $-$  & 0.3 & 12.3 &  0.4 & 12.9 \\
                         & Retrograde Counterpart & Accreted & 1.3 &  13.5  & {\bf 0.2}  & 12.3  & 0.6  & 12.5 & 0.6 & 12.5 &  1.3 & 12.4 \\
                                     & Disk kinematics & In-situ  & 0.3 & 12.6   & 1.0  & 12.9  & $-$  & $-$  & $-$ & $-$  & $-$  & $-$   \\
\hspace{0.7cm} ${\rm [Fe/H]} \leq -$3& Disk kinematics & Accreted & 0.7 & 13.5   & 3.3  & 13.3  & 4.0  & 13.3 & 7.3 & 13.2 & 1.8 & 13.0   \\
                         & Retrograde Counterpart & In-situ      & 0.6 & 13.6   & $-$  & $-$   & $-$  & $-$  & $-$ & $-$  & $-$ & $-$     \\
                         & Retrograde Counterpart & Accreted     & 1.5 & 13.5   & 0.2  & 13.0  & {\bf 1.0}  & 12.8 & 0.9 & 13.2 & 2.7 & 13.2 \\
\hline
\hline
\enddata
%\tablecomments{F = Stellar mass fraction}
%\begin{tablenotes}
%\item \small F = Stellar mass fraction.
%\end{tablenotes}
\end{deluxetable*}

%Xx after the presentation of the results from TNG50
\noindent
An advantage of the simulations is that we can further explore the properties of stellar particles according to their origin, in-situ and accreted, coupled with their rotational characteristics, prograde and retrograde.
Table \ref{Tab:Simulations2} lists the stellar mass fraction and median age for the in-situ and accreted halo stellar components, possessing disk kinematics, and its retrograde counterpart, in the two metallicity intervals. 
%The stellar mass fraction is determined as M$_{\rm F}$ = $\sum_{i} m_{i}$ / %$\sum_{k} m_{k}$, where $i$ and $k$ run within each subsample, and the total sample %in each metallicity range, respectively. 

\noindent
As can be seen in this table, in case of VMP stars with disk kinematics, both Aq-C and TNG50 MW's analogs exhibit a systematic larger contribution of accreted stars. Interestingly, in some of them, there is also a contribution from VMP stars formed in-situ, albeit smaller. Accreted stars are even more dominant in the lower metallicity interval. In fact most of the TNG50 MW analogs do not have an in-situ component at metallicities [Fe/H]$ \leq -$3.  While these results reflect the  diversity of halo properties originated by the particular assembly history of each galaxy, they also suggest a clear trend of a more significant contribution of accreted stars with disk rotation in the VMP regime, as well as the presence of stars originated in-situ.

\noindent
Table \ref{Tab:Simulations2} also shows that in-situ and accreted components are very old with median ages larger than 12.5 Gyr, corresponding to redshift, $z >$ 5. This is consistent with the claim that they have been formed during the very first stages of galaxy assembly (CC21). In the retrograde counterpart the trend is similar with prevalence of accreted, very low metallicity stars. We have also explored the $\alpha$-elements in both simulations, and found that VMP stars with disk kinematics and its retrograde counterpart have high $\alpha$-enrichment, in agreement with their  very old age. We have not noticed any systematic trend in [$\alpha$/Fe] for these groups of stars. 

\noindent
It is important to note that the stellar particles we are investigating are very metal-poor, and very old, in all the analysed halos. Hence, they formed in the first structures that merged to assemble the MW progenitors. Such first structures are expected to be small, and with comparable masses, so the classification of main progenitor, and satellite represents a challenging task. Nevertheless, we have taken all the possible measures to select the main progenitor and follow back in time all stellar particles to their site of origin.

\noindent
While the detailed analysis of the assembly histories will be discussed in a forthcoming paper, it is worth noting that most of the accreted stars come from  small galaxies of $10^{6}-10^{8} \rm M_{\odot}$ and , in some cases, form a  more massive systems of $10^{9} \rm M_{\odot}$. The accretion of these larger systems to the main progenitor galaxy is found to be  around $z \gtrsim 2$.

\vspace{0.5cm}
\subsection{The most bound stars in the halos of simulated galaxies}

\noindent
%{\bf 
%The formation and evolution of stellar halos differ from
%those of dark halos. While the dark halos are entirely formed through a hierarchical
%assembly and merging process \citep{white1978}, stellar components in originate from multiple sources, in particular from star
%formation within cooled gas originally present in parent halos,

Stellar halos are reported to have different channel of formation  such as in gas stripped from merging satellites, or stars supplied by accreted galaxies that
are disrupted by tidal interaction or survive as satellites, or disk-heated stars probably triggered by the accretion of other galaxies
(e.g., \citealt{bekki2001}; \citealt{bullock2005}; \citealt{zolotov2009}; \citealt{purcell2010}; \citealt{font2011}; \citealt{mccarthy2012}; \citealt{tissera2013}; \citealt{cooper2015}; \citealt{deason2016}; \citealt{rodriguez-gomez2016}; \citealt{dsouza2018};
\citealt{monachesi2019}; \citealt{fattahi2020}).
The formation of the inner region of galaxies and the inner part of stellar halo is still an open question because of the difficulty in disentangling the stellar populations coming from different formation channels. In fact, part of these stars could have formed during  a first  collapse of the progenitor, or by the aggregation of small gas-rich clumps, and hence, we expect them to be highly bound. 

%\DCb{What I wrote before makes more sense to me in term of low binding energy. I can see why %stars formed from cooled gas in the bottom of the main progenitor halo, or from %merged/accreted cold gas supplied by other halos are most bound. While, why stars formed from %first collapse of the progenitor, or by the aggregation of small gas-rich clumps would be %most bound? Are we saying the same things?}} 
%cooling, or, in other words, they possess the lowest binding
%energy, E.

%In CC21 (see their discussion in section 4.4.2), where the St\"ackel gravitational potential was adopted, it was shown that metal-poor halo stars with the lowest binding energies, E $< -$1.5 \en, and Lz $\sim$ 0 are the most tightly bound to the Milky Way gravitational potential.}\\
As discussed in Section \ref{Sect:PhaseSpace}, the VMP HES sample contains some of the stars most tightly bound to the Milky Way gravitational potential (E $< -$1.5 \en, and Lz $\sim$ 0 in the adopted St\"{a}ckel - see section 4.4.2 in CC21).
%The VMP HES sample contains such stars as discussed Sect. 3. 
Here, we will search for stellar particles in our simulation suites possessing similar characteristics, and investigate their formation site.

\noindent
By using the gravitational potential of St\"{a}ckel type, we built the ($L_z$, ${\rm E}$) distributions for Aq-C  and the four TNG50 MW's analogs, and selected the stellar particles with
total energy E $< -$1.5 \en. Table \ref{Tab:Simulations3} lists the fraction of VMP most-bound halo HES stars and the corresponding mass fraction for the analyzed halos, in each metallicity range. In case of the HES sample, the fraction is determined by dividing the number of most-bound halo stars by the total number of stars in each range of metallicity. The stellar mass fractions for the simulated galaxies are  
defined as M$_{\rm F}$ = $\sum_{i} m_{i}$ / $\sum_{k} m_{k}$ where $i$
runs within the sub-sample of most bound stars (those with E $< -$1.5
\en), and $k$ runs over the total sample in each metallicity range.  

\noindent
In the HES sample, and in the metallicity range  $-3 < \rm{[Fe/H]} \leq -2.5$, the fraction of the most-bound halo stars is more than twice the same fraction in the lowest metallicity interval ([Fe/H] $\leq -$3), while in the simulated galaxies, the
stellar mass fraction of similarly selected halo stars is comparable in both metallicity ranges, except for the galaxy 517271. 
%It is worth noting that in the extremely
%metal-poor regime ([Fe/H] $<$ $-$3.0), the fraction of stars with E $<
%-$1.5 \en in the HES sample, and the stellar mass fractions in most of the TNG50-MW simulated galaxies are remarkably similar.\\ 

\noindent
Table \ref{Tab:Simulations4} shows the stellar mass fraction, origin (in-situ or accreted), and median age for the most-bound halo stars
in Aq-C, and the four TNG50 MW's analogs, in the two ranges of metallicity. In Aq-C the mass fraction of stellar particles born in-situ is twice than that of the accreted stellar particles, in both metallicity intervals. On the contrary, in case of the TNG50 galaxies, the mass fraction of accreted stellar particles is twice than those with in-situ origin, in both metallicity intervals. 

\noindent
As expected, the VMP most-bound halo stars are very old in both metallicity ranges, with median ages above $\sim 13.3$  Gyr for all the analogs ($z \sim$ 11). They are also $\alpha-$elements enriched. Hence, these stars are among the first stellar populations to be formed in these simulations. Such properties make them an interesting and dynamical informative population, but it also call for caution since at the early stages of galaxy formation, the numerical resolution is limited, even for the TNG50 haloes.
Such limitation is due to the fact that the progenitor subhaloes at $z >5 $ are very small and contain a few number of particles.

\begin{deluxetable*}{lcccccc}
\tabletypesize{\footnotesize}
\singlespace
\tablecolumns{12}
\tabcolsep=0.07cm
\tablewidth{0pt}
\label{Tab:Simulations3}
\tablecaption{The most bound stars in the halo. Fractions of HES stars and stellar mass fractions of simulated galaxies in the two defined metallicity intervals.}
\tablehead{
\colhead{Metallicity} &
\colhead{HES} &
\colhead{Aq-C} &
\colhead{469487} &
\colhead{517271} &
\colhead{509091} &
\colhead{446665} \\
\colhead{} &
\colhead{F (\%)} &
\colhead{M$_{\rm F}$ (\%)} &
\colhead{M$_{\rm F}$ (\%)} &
\colhead{M$_{\rm F}$ (\%)} &
\colhead{M$_{\rm F}$ (\%)} &
\colhead{M$_{\rm F}$ (\%)} \\
}
\startdata 
$-$3 $<$ ${\rm [Fe/H]}$ $\leq -$2.5      & 6.6 & 7.6 & 2.3 & 2.0 & 2.4 & 2.2  \\ 
\hspace{0.7cm} ${\rm [Fe/H]} \leq -$3    & 2.8 & 6.7 & 2.4 & 0.9 & 2.7 & 2.2  \\
\hline
\hline
\enddata
%\tablecomments{}
\end{deluxetable*}

\begin{deluxetable*}{llccccccccccc}
\tabletypesize{\footnotesize}
\singlespace
\tablecolumns{12}
\tabcolsep=0.07cm
\tablewidth{0pt}
\label{Tab:Simulations4}
\tablecaption{The most bound stellar populations in the halo:  Mass fractions of in-situ and accreted stars (given in percentages), and their median age (in Gyr) for the analyzed simulated halos. Boldface numbers denote mass fractions within three bootstrap $\sigma$}.
\tablehead{
\colhead{Metallicity} &
\colhead{Origin} &
\colhead{} &
\colhead{Aq-C} &
\colhead{Age} &
\colhead{469487} &
\colhead{Age} &
\colhead{517271} &
\colhead{Age} &
\colhead{509091} &
\colhead{Age} &
\colhead{446665} &
\colhead{Age}\\
\colhead{} &
\colhead{} &
\colhead{} &
\colhead{M$_{\rm F}$ (\%)} &
\colhead{Gyr} &
\colhead{M$_{\rm F}$ (\%)} &
\colhead{Gyr} &
\colhead{M$_{\rm F}$ (\%)} &
\colhead{Gyr} &
\colhead{M$_{\rm F}$ (\%)} &
\colhead{Gyr} &
\colhead{M$_{\rm F}$ (\%)} &
\colhead{Gyr}
}
\startdata 
$-$3 $<$ ${\rm [Fe/H]}$ $\leq -$2.5    & In-situ     && 1.2  & 13.7 & {\bf 1.0} & 13.4 & 0.1 & 13.1  & 0.5 & 13.0 & 0.4 & 13.2  \\ 
                                       & Accreted    && 6.4  & 13.5 & {\bf 1.3} & 13.2 & 1.9 & 13.3 & 1.9 & 13.4 & 1.8 & 13.3  \\
\hspace{0.7cm} ${\rm [Fe/H]} \leq -$3  & In-situ     && 1.4  & 13.7 & 0.5 & 13.6 & $-$ & $-$  & $-$ & $-$  & 0.5 & 13.7  \\
                                       & Accreted    && 5.3  & 13.6 & 1.9 & 13.6 & 0.9 & 13.3 & 2.7 & 13.3 & 1.7 & 13.5 \\
\hline
\hline
\enddata
%\tablecomments{}
\end{deluxetable*}

%{\bf Patricia I moved the following paragraphs from the introduction\\}
\subsection{Discussion}
\noindent
In the current %$\Lambda$ Cold Dark Matter (LCDM) 
$\Lambda$CDM cosmological paradigm, the stellar halos of galaxies are expected to be formed mainly by the accretion of satellite galaxies  with a smaller contribution of in-situ stars in the inner region \citep{zolotov2009,cooper2015,font2011,tissera2012}.
It has been shown that few more massive satellites (i.e., a few times $10^{9}$M$\odot$)  could have contributed to the formation of inner halo while the outer halo was likely formed by larger contributions from smaller ones \citep{tissera2014,monachesi2019,fattahi2020}. Most of these investigations have been focused on the study of MW mass-sized galaxies and numerous works have reported the existence of stellar streams and remnants of the satellites accretion \citep{helmi2008}. 
Numerical simulations of MW mass-sized galaxies provide the opportunity to reconstruct the assembly history of the stellar halo as well as their host galaxies \citep{mackereth2019,bignone2019} and hence, interpret and even predict the possible formation sites of stellar populations with particular characteristics, such as the most-bound halo stars, and the very metal-poor stars with disk kinematics  found in different observational surveys \citep{sestito2019,sestito2020,dimatteo2020,venn2020}, and in this paper. \\

\noindent
\citet{sestito2021} analysed the properties of stellar particles selected to be VMP and UMP in a suite of  NIHAO-UHD simulations, mimicking the observational selection of \citet{sestito2019} and \citet{sestito2020}. They found that a fraction of VMP and UMP halo stars coexist with the disc component. These stars have prograde or retrograde co-planar orbits consistent with the observations. Because of their very low [Fe/H], the star particles are found to be old with ages larger than 12.5\,Gyr.

\noindent
The advantage of exploring simulations to shed light on the origin of these old, VMP stars is given by the possibility of following star particles back in time to identify their formation site. Following this approach, \citet{sestito2021} envisaged two possible scenarios to explain the origin of VMP stars with disk kinematics. The first, which seems to be the dominating one in these simulations, is consistent with the assembly of the first protogalaxy via a chaotic process of merging and accretion of small systems of about $10^{5-9}\rm{M_{\odot}}$. The proto-galaxy and the proto-disk are assembling in a gravitational potential shallower than the present day galaxy, which causes the small merging systems to deposit star particles in the inner region of the main halo, in prograde and retrograde motion. \\

\noindent
The second scenario involves a later merger event whose stars are stripped and assimilated in the disc components depending on their orbital parameters. A third scenario which assumes that the VMP stars could have been formed from low-metallicity regions in the disks was discarded in the NIHAO-UHD simulations. This was based on the fact that all VMP star particles in the simulated halos are born $\approx 4$~Gyr before the formation of a stable disk.

\noindent
The findings reported by \citet{sestito2021} are also in global agreement with previous studies of the contribution to the disk component  by satellites accreted at later stages of disc formation \citep{abadi2003,tissera2012}, and these accretion could contribute with old, very metal poor stars \citep{scannapieco2011,gomez2017}. The retrograde coplanar components are more likely to be formed in the first scenario which proposes  a violent process of assembly.\\

\noindent
The simulations explored in this paper, Aq-C, 
%and the merger trees, properties of the progenitors, and accreted satellites of 
and TNG50 MW'analogs have shown that most of the VMP star particles with disk kinematics, and 
its retrograde counterparts, are accreted. The contributing satellites are within the range of $[10^6-10^9]$~M$_\odot$, and are very gas-rich, with %gas-and-stellar 
gas-to-stellar mass ratios larger than  unity.
Such findings are in agreement with the first scenario envisaged by \citet{sestito2021}. 
However, our analysis shows that a fraction of VMP stars with disk kinematics was formed in-situ, in rotational supported structures present in the progenitor galaxies. Such a possibility was discarded by \citet{sestito2021}, while Aq-C and TNG50 MW analogs show that this could be a possible channel of formation for a fraction of such stars (see Table \ref{Tab:Simulations2}).  We acknowledge the fact that, even within the TNG50 simulated galaxies, these results should be taken as indicative because of the limitation on the numerical resolution at very high redshift. A more detailed analysis of these VMPs in the TNG50 MW's analogs will be presented in a forthcoming paper (Sillero et al., in preparation).\\

\noindent
As suggested by CC21, VMP stars with disk kinematics were likely formed by
early infalling, pristine gas settled into an
equatorial plane of a progenitor dark halo in the presence of an
initial angular momentum (e.g., \citealt{katz1991}). From this
gas, a fraction of very metal-poor stars, with low \I and
high $L_z$, may have been formed and remained in the disk. Recent works in the extra-galactic domain show the existence of gaseous disks in star forming galaxies at high redshift. For example, \citet{neeleman2019} and \citet{neeleman2020} used data from the Atacama Large Millimeter/submillimeter Array where galaxies were selected through their [C II] emission spectroscopy of H I absorption, and found gaseous disks at redshifts z $\sim$ 4-5 (12 Gyr ago), which is consistent with the median age found for VMP stars with disk kinematics in the MW's analogs. The existence of these gaseous disks supports a scenario in which cold gas would be accreted onto dark halos during the early stages of the galaxy assembly \citep{dekel2009}. Within these 
%highly
rotating primordial disks, star formation would take place forging very metal-poor stars with disk kinematics, such as those observed in this analysis and other works mentioned earlier in the paper. The properties and origins of these stars suggest that they could be considered what remain of the primordial disk formed during the initial stages of galaxy assembly at very high redshift.

\section{Summary}

\noindent
In this paper we explored the chemo-dynamical properties of a stellar sample obtained by matching metal-poor stars selected from HES with Gaia EDR3. The survey selected primarily very metal-poor stars, and the adopted sample has a completeness of 50\% at [Fe/H] $= -2.5$. The analysis is based on the phase-space defined by the three integrals of motion (E, $L_z$, $I_3$), combined with the metallicity. $I_3$ is the third integral of motion derived by using the analytical definition in the St\"{a}ckel gravitational potential. In the analysis, we considered only the average properties, or {\it coarse-grained} phase-space distribution of the halo. The main findings can be summarized as follows:

\begin{itemize}
\item The $L_z$ distribution of the very metal-poor halo appears truncated in the prograde high-$Lz$ side, at $L_z \sim $ 1500 \kpckms, while the negative side shows an asymmetric distribution in $L_z$ towards large retrograde motion and high energy. At [Fe/H] $< -$2, most of the stars possess \I $>$ 500 \kpckms (\tho $>$ 5-6 deg) and $L_z <$ 1500 \kpckms. This results are in agreement with what found in CC21.

\item Some stars with metallicity [Fe/H] $< -$2 have disk kinematics, having $L_z >$ 1500 \kpckms, and \I $<$ 1000 \kpckms. We found 33, 13, 9, 2, and 2 stars in the metallicity intervals of $-$2.5 $<$ [Fe/H] $< -$2.0, $-$3.0 $<$ [Fe/H] $< -$2.5, $-$3.5 $<$ [Fe/H] $< -$3.0, $-$4.0 $<$ [Fe/H] $< -$3.5, and [Fe/H] $< -$4.0, respectively. 

\item In the HES sample the number of stars with [Fe/H] $< -$2.5 and disk kinematics is more than twice as large as its counterpart with retrograde motion.

\item At [Fe/H] $< -$2.5 we found 37 halo stars with the lowest binding energy ($E < -$1.5 \en), and $L_z \sim$ 0 \kpckms. These are stars most tightly bound to the gravitational potential of the main progenitor halo. %likely formed from cooled gas in the bottom of it, or from merged/accreted cold gas supplied by other halos, and can be considered candidate in-situ halo stars. 
We found 31, 5, and 1 of such stars in the metallicity intervals of  $-$3.0 $<$ [Fe/H] $< -$2.5, $-$3.5 $<$ [Fe/H] $< -$3.0, and [Fe/H] $< -$4.0, respectively. 

\item The majority of stars in the very metal-poor halo have orbital angles in the range of 5--7\,deg $<$ \tho $<$ 30--40\,deg (80\%), with the exception of those with disk kinematics falling in the range of 1--2\,deg $<$ \tho $<$ 12\,deg. The retrograde counterpart of the VMP sample with disk kinematics appears to be less planar having orbital angles \tho $>$ 20--30\,deg. 

\item  We identified VMP stars with disc kinematics in the halos of five MW analogs, Aq-C and four TNG50 galaxies. When followed back in time along their assembly histories, we found that these stars had two different origins: 1) accretion from early satellite galaxies, which is the dominant mechanism of formation, and 2) in-situ formation, particularly in the metallicity interval of $-3 <\rm [Fe/H] < -$2.5. Both type of stars are very old, with ages $> $12.5 Gyr, and $\alpha$-enriched. The retrograde counterparts exhibit also two mechanisms of formation, accreted and in-situ, with dominance of accretion. For the TNG50 analogs, which can be followed back in time with higher numerical resolutions, we identified contributing gas-rich satellites with stellar mass within $[10^6, 10^9]\, \rm M_{\sun}$. Further analyses focused on the origin of VMP stars with disk kinematics and its retrograde counterpart will be carried out in a separate paper (Sillero et al., in preparation).

\item All simulated halos analysed in this work have stellar populations which resemble the most bound VMP halo stars reported in the observations. These stars are the oldest detected, with a median age $\sim$ 13.3\,Gyr and $\alpha$ enriched. While we could, in principle classify them as accreted and in-situ, it is not possible to draw a robust conclusions on their formation, since they are amongst the first stars formed in the inner region of halos and, hence, are resolved by very few particles. However, it is clear that these very old, most-bound VMP stars contain important information about the first stages of the formation of our Galaxy and the first stars formed in the Universe.
\end{itemize}

\section*{Acknowledgments}

\noindent
We would like to thank Annalisa Pillepich for her valuable suggestions and help with TNG50 data and Timothy Beers for his work on the derivation of the metallicities.
We would also like to thank our colleagues who helped with acquiring medium-resolution spectroscopy of the metal-poor HES stars used in this paper: Paul Barklem, Berit Behnke, Michael S. Bessell, Paul Cass, Judith L. Cohen, Cora Fechner, Anna Frebel, Birgit Fuhrmeister, Malcom Hartley, Dionne Haynes, Andrew McWilliam, Jorge Melendez, John E. Norris, Ivan Ramírez, Ken Russell, Stephen Shectman, Ian Thompson, Fred Watson, and Franz-Josef Zickgraf.

\noindent
NC acknowledges funding by the Deutsche Forschungsgemeinschaft (DFG, German Research Foundation) -- Project-ID 138713538 -- SFB 881 (``The Milky Way System'', subproject A04).

\noindent
PBT acknowledges partial funding from Fondecyt 2021/1200703 (ANID), and CATA-Basal- FB210003 and Nucleo Milenio ANID ERIS.  We acknowledge the use of  Ladgerda Cluster funded by Fondecyt 2021/1200703 (ANID).
% !!!!
% Please use BibTeX! It makes it MUCH easier to handle references.
% !!!
{}

\end{document}